\documentclass[sn-basic]{sn-jnl}% Basic Springer Nature Reference Style/Chemistry Reference Style

\usepackage{graphicx}%
\usepackage{multirow}%
\usepackage{amsmath,amssymb,amsfonts}%
\usepackage{amsthm}%
\usepackage{mathrsfs}%
\usepackage[title]{appendix}%
\usepackage{xcolor}%
\usepackage{textcomp}%
\usepackage{manyfoot}%
\usepackage{booktabs}%
\usepackage{algorithm}%
\usepackage{algorithmicx}%
\usepackage{algpseudocode}%
\usepackage{listings}%
\usepackage{svg}
\theoremstyle{thmstyleone}%
\theoremstyle{thmstyletwo}%

\theoremstyle{thmstylethree}%

\newcommand{\MgII}{Mg~\textsc{ii}}
\newcommand{\FeII}{Fe~\textsc{ii}}
\newcommand{\SiII}{Si~\textsc{ii}}
\newcommand{\OI}{O~\textsc{i}}
\newcommand{\NaD}{Na~\textsc{i~d}}
\newcommand{\kms}{\mbox{km s$^{-1}$}}
      
\begin{document}

\title[Galactic Outflows at Cosmic Noon]{Galactic Outflows and Star-Formation Quenching: A JWST Perspective}


\author*[1]{\fnm{Rebecca L.} \sur{Davies}}\email{rdavies@swin.edu.au}
\author*[2]{\fnm{Sirio} \sur{Belli}}\email{sirio.belli@unibo.it}

\affil[1]{\orgdiv{Centre for Astrophysics and Supercomputing}, \orgname{Swinburne University of Technology}, \orgaddress{\street{John Street}, \city{Hawthorn}, \postcode{3122}, \state{VIC}, \country{Australia}}}
\affil[2]{\orgdiv{Dipartimento di Fisica e Astronomia}, \orgname{Universit\`a di Bologna}, \orgaddress{\city{Bologna}, \country{Italy}}}


\abstract{Outflows driven by active galactic nuclei (AGN) are widely considered a key mechanism for regulating and quenching star formation in massive galaxies. However, direct observational evidence linking outflows to quenching has remained elusive for decades. This article reviews recent findings for massive galaxies at cosmic noon ($z\sim 2$), based on sensitive James Webb Space Telescope (JWST) observations tracing both star formation histories (through Balmer absorption lines) and neutral outflows (through blueshifted \NaD\ absorption). Neutral outflows are found in $\sim$~25~--~50\% of massive ($\log(M_*/M_\odot)>$~10) galaxies, independent of star formation rate. Outflows from star-forming galaxies are consistent with being powered by star-formation and/or AGN activity, whereas outflows from quiescent systems are generally attributed to current or previous AGN activity. There is growing evidence that recent rapid quenching of star-formation is associated with powerful AGN-driven outflows with mass loading factors \mbox{$\eta \gtrsim$~1~--~10}. The neutral gas dominates the outflow mass budget but in most cases its velocity is below the halo escape velocity, indicating that this gas may form part of a galactic fountain. Detections of neutral gas inflows and residual cold gas reservoirs in quiescent galaxies suggest that other feedback mechanisms such as AGN heating are required to keep galaxies quenched over long timescales. We highlight the potential of multiphase outflow measurements to constrain feedback models, describe the main systematic uncertainties in mass outflow rate measurements, and discuss how the $z\sim 2$ results inform our understanding of the formation of the earliest massive quiescent galaxies.}


\maketitle

\section{Introduction}\label{sec1}

Understanding what causes massive galaxies to quench is one of the biggest outstanding problems in modern astrophysics (see \citealt{Bluck26a, Bluck26b, Whitaker26} for reviews). Cosmological hydrodynamical simulations (e.g. Illustris-TNG; \citealt{Pillepich18}, EAGLE; \citealt{Schaye15}, SIMBA; \citealt{Dave19}, Magneticum; \citealt{Steinborn15}) generally require some form of feedback from active galactic nuclei (AGN) to reproduce the observed galaxy stellar mass function, star-formation rate distribution and color bimodality \citep[see also][]{Benson03, DiMatteo05, Springel05, Bower06, Croton06, Hopkins06, Somerville08, Erb15, Beckmann17}. However, the exact physical processes coupling AGN accretion energy to the host galaxy remain a matter of debate. Feedback processes are often divided into two broad categories: ejective or quasar-mode feedback, which expels gas from galaxies through outflows, and preventative or radio-mode feedback, which heats gas, preventing it from cooling and being re-accreted to replenish the reservoir of fuel for star-formation (see \citealt{Harrison24} and references therein). The state-of-the-art simulations mentioned above employ vastly differing combinations of ejective and preventative feedback and yet, because of the large number of free parameters that can be tuned, they produce galaxies with roughly the same stellar properties \citep[e.g.][]{Crain23, Wright24}. 

Observational studies have taken a variety of approaches to constrain the primary driver(s) of quenching. For example, large statistical studies of galaxies spanning $z\sim$~0~--~2.5 have investigated which properties most strongly predict whether a galaxy is quenched. While stellar mass is a good predictor \citep{Peng10}, central density, velocity dispersion, or bulge mass are even better predictors \citep[e.g.][]{Franx08, Lang14, Bluck22}. This result is often interpreted as a sign that long-term AGN heating of halo gas may be the primary quenching mechanism. However, directly constraining the timescale and causes of quenching is substantially more challenging. Most local quiescent galaxies quenched $\sim$~8~--~10~Gyr ago \citep[e.g.][]{Ilbert13}, but spectroscopic signatures of rapid quenching and the powerful outflows that may trigger it disappear after $\sim$~1~Gyr \citep[e.g.][]{Park24, Sun24, Taylor26}. Therefore, to robustly identify the physical processes that quench massive galaxies, it is necessary to study galaxies that are currently undergoing, or recently underwent, quenching. 

To this end, studies of post-starburst galaxies have provided important insights into galaxy quenching mechanisms. Post-starburst galaxies are distinguished by weak ionized gas emission lines and strong Balmer absorption lines, indicating a dominant population of A-stars \citep{Dressler83}. The prevalence of A stars combined with the absence of O and B stars suggests that these galaxies must have experienced a strong burst of star-formation a few hundred Myr prior to the epoch of observation, followed by rapid quenching. 

Post-starburst galaxies are rare at $z\sim$~0, accounting for $<$~1\% of all massive galaxies \citep[e.g.][]{Zabludoff96, Quintero04, Goto07, Alatalo16, Chen19}. Prior to JWST, there were very few high-quality Balmer absorption line spectra for galaxies beyond the local Universe, driving the development of alternate methods to identify post-starburst galaxies, primarily using combinations of rest-frame colors \citep[e.g.][]{Wild14, Belli19}. Number densities calculated using these methods suggest post-starburst galaxies are more common at $z\sim$~2, accounting for $\sim$~5\% of all massive galaxies \citep[e.g.][]{Wild16}. Furthermore, as many as half of all galaxies entering the red sequence at $z\sim$~2 could go through the post-starburst phase \citep[e.g.][]{Forrest18, Rowlands18, Belli19, DEugenio20, Wild20, French21, Skarbinski26}. The other half likely quench over longer timescales ($\sim$~1~Gyr) comparable to the typical gas depletion times at these redshifts \citep[e.g.][]{Tacconi20}. This slow quenching is thought to be consistent with starvation, i.e. the cessation of gas replenishment due to inflows \citep[e.g.,][]{Peng15}.

High-velocity outflows are found in approximately half of all massive post-starburst galaxies at $z\sim$~0 \citep[e.g.][]{Davis12, Alatalo16, Baron20, Baron22}, and these outflows appear to decline in strength and velocity with time since quenching \citep[e.g.][]{Sun24}, suggesting that the sudden quenching is driven by rapid ejection of cold gas. However, prior to JWST, there were very few direct constraints on the mass outflow rates from post-starburst galaxies outside of the local Universe. Early studies of post-starburst galaxies at $0.5 \lesssim z \lesssim 2$ were based on small samples and/or stacked spectra and revealed tentative evidence for widespread, high velocity outflows similar to what is observed at $z\sim$~0 \citep[e.g.][]{Tremonti07, Maltby19, Man21, Taylor24}. Deep ALMA observations of post-starburst galaxies at local and intermediate redshift ($z\lesssim$~0.6) have revealed that their molecular gas reservoirs also decline rapidly with time after quenching, at a rate faster than can be explained by the residual star-formation activity in these systems, providing further evidence for significant ejection of molecular gas by outflows \citep[e.g.][]{French18, Bezanson22, Baron23}. Measuring molecular gas reservoirs in more distant post-starburst galaxies is very challenging even with sensitive sub-mm telescopes like ALMA and NOEMA. Initial observational results based on small samples support the $z\sim$~0 picture that gas reservoirs decline rapidly with time after quenching, but there is substantial diversity in the amount of molecular gas measured for different samples \citep[e.g.][]{Williams21, Belli21, Whitaker21, Zanella23, Suess25}. Although there is compelling circumstantial evidence for a link between outflows and rapid quenching, robust measurements of outflow rates and molecular gas reservoirs for post-starburst galaxies at cosmic noon are needed to verify this directly.

The unprecedented near-infrared sensitivity of JWST has transformed our understanding of AGN-driven outflows and their link with quenching in massive galaxies at cosmic noon. This review outlines those advances, with a specific focus on insights gained from the first observations of neutral outflows in representative samples of galaxies beyond the local Universe. 

\section{Situation prior to JWST}

\subsection{Multiphase gas in galaxies and outflows}\label{subsec:multiphase_properties}
The interstellar medium (ISM) is multiphase, containing cold molecular gas (\mbox{$T\sim$~10~--~100~K}, $n\sim$~10$^{3-6}$ cm$^{-3}$), cool neutral atomic gas ($\sim$~100~K, $\sim$~30 cm$^{-3}$), warm neutral and ionized gas ($\sim$~10$^4$~K, $\sim$~0.1~--~1~cm$^{-3}$), and hot diffuse gas ($\sim$10$^{5-6}$~K, $\sim$~10$^{-3}$ cm$^{-3}$) \citep[e.g.][]{Field69, McKee77, Wolfire95, Cox05, Saintonge22}. The hot and warm ionized phases have low densities but large filling factors, and in the Milky Way these phases account for $\sim$~20~--~25\% of the total ISM mass \citep[e.g.][]{Ferriere01, Girichidis20}. The remaining 75\% of the ISM mass is found in the denser neutral and molecular phases \citep{Draine11, Saintonge22}. The neutral phase is dominant at $z\sim$~0 \citep[e.g.][]{Boselli14, Catinella18}, but the molecular phase is expected to account for a larger fraction of the mass at higher redshifts due to the conversion of H~\textsc{i} to H$_2$ at high gas mass surface densities \citep[e.g.][]{Tacconi20}.

Outflows comprise a combination of hot wind ($\sim$10$^{6}$~K) material (e.g. supernova ejecta or material launched from black hole accretion disks) and gas swept-up from the ISM \citep[e.g.][]{King15, Bieri17, Richings18, Fielding22}. Cool gas may also precipitate out of the hot wind due to radiative cooling in turbulent mixing layers \citep[e.g.][]{Gronke18, Ji19, Fielding20, Chen23}. As a result, outflows consist of multi-phase gas spanning many orders of magnitude in temperature and density \citep[e.g.][]{Rupke05b, Bolatto13, Cicone14, Feruglio15, Veilleux20}. Disk-scale ISM simulations of both supernova-driven winds (e.g. CGOLS; \citealt{Schneider18}, TIGRESS; \citealt{Kim20}, QUOKKA; \citealt{Vijayan24}) and AGN-driven outflows \citep[e.g.][]{Ward24} predict that similar to the ISM, the majority of the mass ejected in outflows is found in the cooler neutral and molecular phases. Consistent with this prediction, multiphase studies of outflows in the local Universe generally find that ionized gas outflow rates are comparable to or lower than atomic and/or molecular mass outflow rates, although there is significant variation in measured ratios between studies \citep[e.g.][]{Fiore17, Oosterloo17, Fluetsch19, Avery22, MazzilliCiraulo25, HernandezYevenes26}.

\subsection{Outflows at Cosmic Noon}\label{subsec:cosmic_noon}
The detection of prominent blueshifted rest-frame UV absorption lines in stacked spectra of Lyman Break Galaxies provided one of the earliest indications that outflows with velocities of several hundred \kms\ are widespread in star-forming galaxies at cosmic noon \citep[e.g.][]{Pettini01, Adelberger03, Shapley03, Steidel10}. The next key breakthroughs came from near-infrared integral field spectroscopy studies which enabled detailed measurements of outflow properties. Observations of luminous AGN (quasars, bright radio galaxies, and AGN-hosting ultraluminous infrared galaxies) revealed galaxy-scale outflows extending 5~--~10~kpc from the nuclei. These outflows were estimated to release a comparable amount of energy over their lifetimes to the gravitational binding energy of their host galaxies, establishing AGN-driven outflows as a potentially important mechanism for quenching star-formation at cosmic noon \citep[e.g.][]{Nesvadba08, Alexander10, Harrison12, Brusa15, Zakamska16}. Some works reported evidence for localised suppression of star-formation in the paths of powerful outflows \citep[e.g.][]{CanoDiaz12, Cresci15, Carniani16}, although subsequent studies have shown that these conclusions are very sensitive to assumptions regarding the dominant ionization mechanism and the spatial distribution of dust across the systems \citep[e.g.][]{Scholtz21, Kakkad23}.

A key question raised by these studies was whether powerful outflows are only driven by the most luminous AGN. Over the past decade, near-infrared galaxy surveys have revealed that $>$~50\% of massive ($\log(M_*/M_\odot)> 11$) galaxies at cosmic noon show ionized outflows whose properties are most consistent with being AGN-driven \citep[see][and references therein]{NMFS20}. These outflows have typical velocities of \mbox{1000~--~2000~\kms}, exceeding the halo escape velocities and enabling long-term removal of star-forming gas \citep[e.g.][]{NMFS14, Genzel14, Harrison16, Leung19, NMFS19}. Most ionized outflow emission originates from the central 2~--~3~kpc \citep[e.g.][]{NMFS14, Genzel14}, although fainter high velocity emission is detected to $\gtrsim$~5~kpc \citep[e.g.][]{Davies20}; consistent with the outflow extents measured for higher luminosity AGN \citep[e.g.][]{Harrison12, Cresci15}. These high-velocity nuclear outflows have distinctly different properties from star-formation driven winds, which are most prevalent in galaxies above the main sequence \citep[e.g.][]{NMFS19} and originate from kiloparsec-scale star-forming clumps that are often offset from galaxy nuclei \citep[e.g.][]{Genzel11, Newman12_406690}. Star-formation driven outflows have typical velocities of \mbox{300~--~600~\kms}; below the escape velocities of their host galaxies, suggesting that most material launched in star-formation-driven outflows will eventually return to the galaxies in galactic fountains \citep[e.g.][]{Davies19, NMFS19, Freeman19}. Resolved observations suggest that outflows from individual star-forming clumps may propagate $\sim$~1~kpc from their launch point \citep[e.g.][]{Newman12_406690}, with most star-formation-driven outflow emission confined within the effective radius of the host galaxies \citep[e.g.][]{Newman12_global}.

Despite the high incidence of fast outflows in typical massive galaxies at $z\sim$~2, the mass outflow rates measured from ionized gas emission line tracers are generally lower than the star-formation rates of the host galaxies, in contrast with findings for more luminous AGN \citep[e.g.][]{Fiore17} and predictions from galaxy formation models \citep[e.g.,][]{Nelson19, Mitchell20}. These results suggest that although outflows are a prominent feature of typical massive galaxies at cosmic noon, they may not have a major impact on the cycle of gas and are therefore unlikely to be a major cause of quenching \citep[e.g.][]{NMFS19, Concas22}. However, as discussed in Section \ref{subsec:multiphase_properties}, ionized gas likely traces only a small fraction of the outflowing material. Thus, measurements of the cool gas mass carried by distant outflows are crucial to understand their impact on molecular gas reservoirs and their role in galaxy quenching.

Low-ionization absorption lines in rest-frame UV spectra (e.g. \MgII, \FeII, \SiII, \OI) have ionization potentials close to hydrogen and thus trace neutral or partially ionized gas. Although blueshifted rest-frame UV absorption lines have been detected in many star-forming galaxies at cosmic noon \citep[e.g.][]{Weiner09, Steidel10, Kornei12, Erb15}, very few studies report mass outflow rates measured from these lines because 1) the fraction of the total outflowing mass probed by any tracer of partially ionized gas has a large associated uncertainty \citep[e.g.][]{Chisholm16} and 2) the optical depth is degenerate with the small-scale covering fraction of gas against the background continuum source. This degeneracy can be broken using doublet features such as \MgII\ for which the equivalent width ratio of the lines varies with optical depth. However, \MgII\ absorption is often saturated, providing only a lower limit on the outflowing mass. Another key limitation of rest-frame UV absorption line tracers is that they require a bright rest-frame UV continuum, and thus do not provide a viable means of probing outflows in very dusty galaxies or quenching systems \citep[e.g.][]{Kehoe24}.

Most measurements of molecular outflows at cosmic noon prior to JWST came from sub-millimeter observations of CO emission lines in quasars \citep[e.g.][]{Carniani17, Vayner17, Brusa18, Chartas20, Vayner21, Zhong23}, although some possible detections have also been reported in main sequence galaxies \citep[e.g.][]{HerreraCamus19}. Deep stacking analyses using large samples of main sequence galaxies have yielded non-detections of CO outflows down to stringent limits \citep[e.g.][]{Barfety25, Langan26}. These CO outflow non-detections do not preclude the presence of a dominant molecular gas component in outflows from typical star-forming galaxies; this gas may be heated by AGN ionizing radiation, suppressing the CO emission and leading to brighter emission in warm molecular gas tracers such as mid-infrared $H_2$ lines \citep[e.g.][]{DaviesRic24, FernandezOntiveros26}. 

\begin{figure}
\centering
\includegraphics[width=\textwidth, clip = True, trim = 0 0 0 0]{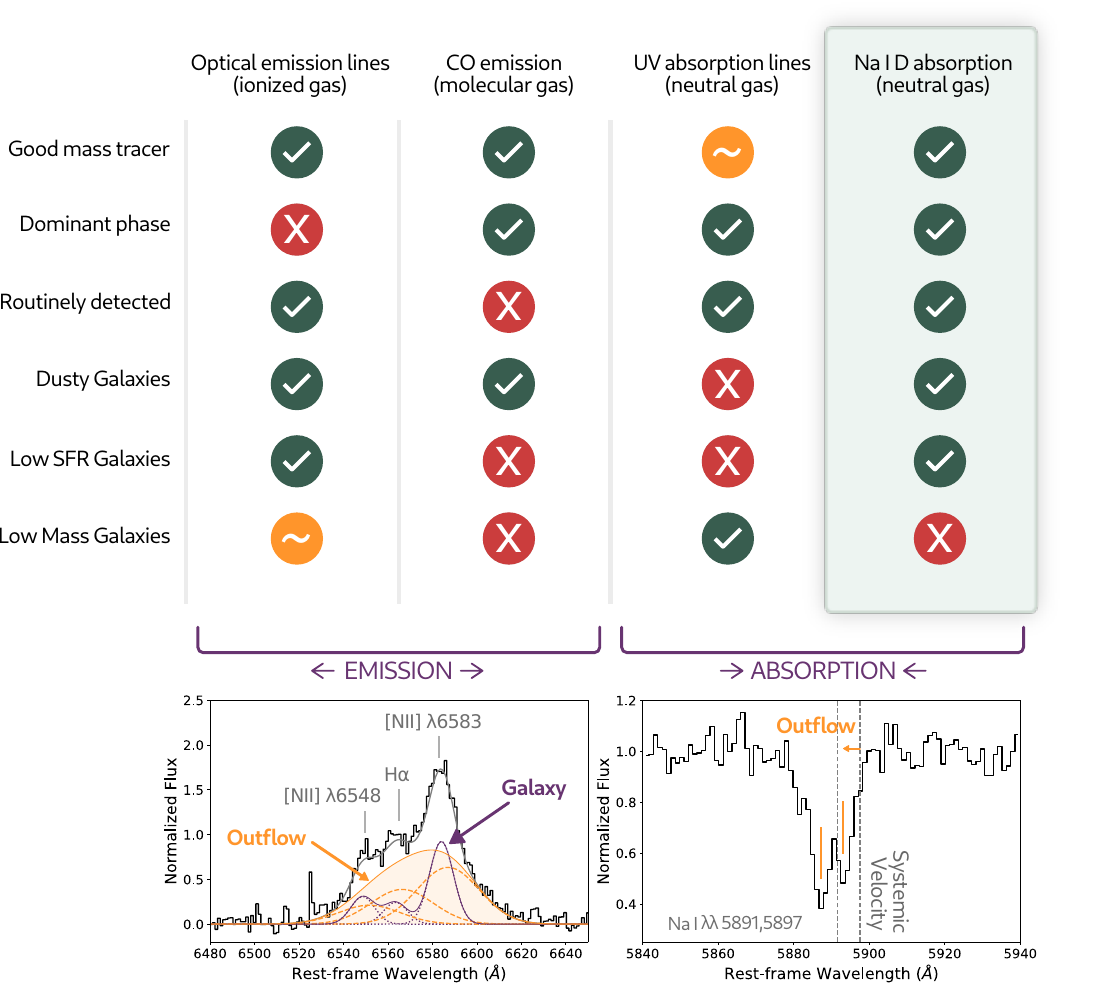} 
\vspace{-10pt}
\caption{Strengths and weaknesses of outflow tracers commonly used at cosmic noon (top). The differences between emission line and absorption line tracers are illustrated using zoom-ins of an example $z\sim$~2 galaxy spectrum (bottom; Davies et al., in prep). Broad optical emission line components tracing ionized outflows are detected in many galaxies at cosmic noon, but they do not trace the dominant outflow phase and often are not sensitive to outflows in low mass galaxies because they blend with the much stronger ISM emission. CO outflows are normally only detected in the brightest systems. UV absorption lines require a bright rest-frame UV continuum and generally cannot be used to probe outflows from dusty or quenching galaxies. \NaD\ absorption is the only tracer that is routinely detected even in dusty and low SFR galaxies and is a good tracer of the mass in the dominant outflow phase. However, it is generally not detected in low mass galaxies, likely due to insufficient dust shielding. Bottom left: Emission-line outflows manifest as broad components (orange) super-imposed on narrower emission from the host galaxy (purple). Dotted and dashed curves show Gaussian profiles of individual emission lines, and solid curves show the summed profile of each component. Bottom right: Absorption-line outflows manifest as blueshifted components relative to the galaxy systemic velocity.} \label{fig:tracers}
\end{figure}

Fig.~\ref{fig:tracers} summarizes the advantages and disadvantages of the main observational tracers used to study outflows from massive galaxies at cosmic noon.

\subsection{Na~I~D as a neutral outflow tracer}
In the local Universe, a common tracer of neutral outflows is \NaD, i.e., the Na~I~$\lambda \lambda$5891, 5897 doublet absorption \citep[e.g.][]{Heckman00, Chen10, Concas19, RobertsBorsani19, Avery22, Weng26}. Neutral sodium has an ionization potential of 5.1~eV -- significantly lower than hydrogen -- and therefore exists primarily in neutral regions where it is shielded by significant columns of gas and dust \citep[e.g.][]{Savage96, Baron20}. \NaD\ absorption is prominent in the spectra of late-type stars and therefore the stellar contribution must be fitted and removed prior to modelling the interstellar contribution. The sodium doublet is resonant, meaning that the interstellar component can appear in either emission or absorption. In the local Universe, galaxies with stellar masses below $\log(M_*/M_\odot)\simeq$~10 typically show \NaD\ in emission. Massive, low SFR galaxies show weak absorption at the galaxy systemic velocity, which is typically attributed to excess stellar absorption. The most massive, highly star-forming galaxies show blueshifted \NaD\ absorption, tracing neutral outflows \citep[e.g.][]{Chen10, Concas19, RobertsBorsani19, Avery22}. 

\NaD\ mitigates many of the issues associated with other outflow tracers at cosmic noon (Fig.~\ref{fig:tracers}). Unlike rest-frame UV lines, \NaD\ absorption is routinely detected in dusty and low SFR galaxies, and it is typically not saturated, providing robust constraints on optical depth and covering fraction \citep{Chen10, Weng26}. The main limitation of \NaD\ absorption is that it is typically undetected in low mass galaxies, likely due to insufficient dust shielding in lower metallicity environments \citep[e.g.][]{RobertsBorsani19}. Unfortunately, at $z\gtrsim$~0.7, \NaD\ is shifted to the near-infrared, where contamination from the Earth's atmosphere makes it very difficult for ground-based telescopes to reach the sensitivity required to robustly detect stellar continuum emission at high signal-to-noise ratio. 

\section{Advances from JWST: Neutral outflows and their connection to rapid quenching}
\begin{figure*}
\centering
\includegraphics[width=\linewidth]{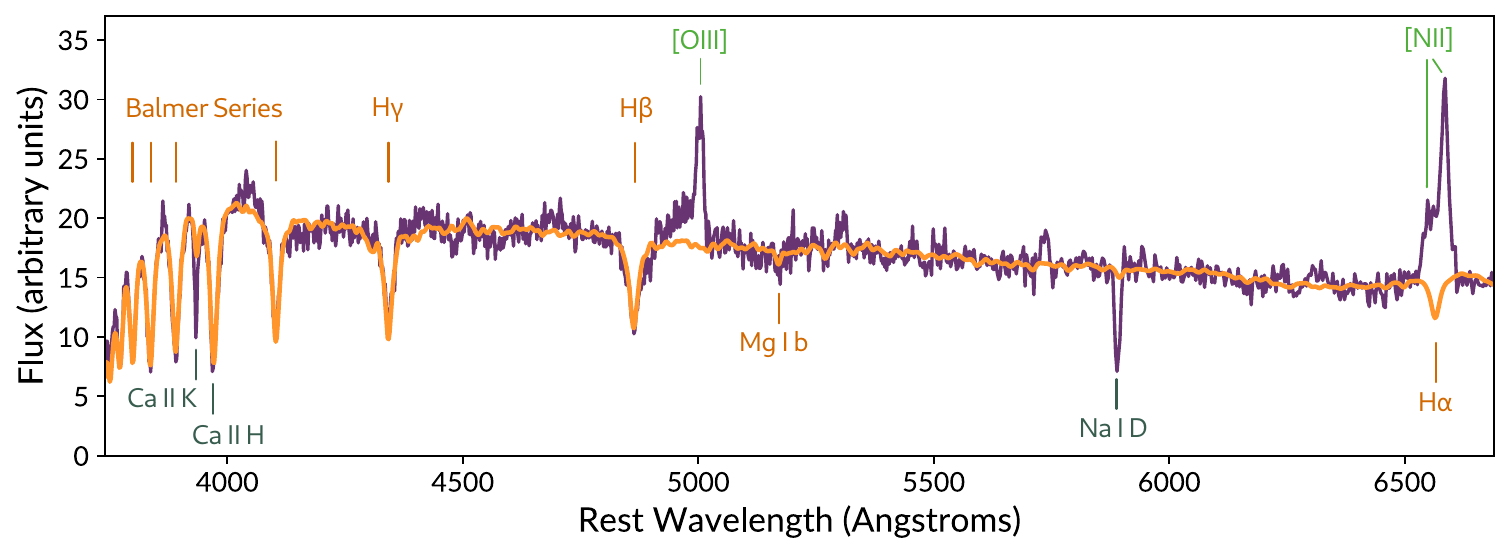} 
\caption{JWST/NIRSpec medium-resolution (R~$\simeq$~1000) spectrum of galaxy COSMOS-11142 at $z$~=~2.445 from GO 1810 \citep{Belli25}. The data is shown in purple and the best-fit \textsc{Prospector} stellar continuum model is shown in orange. Key emission lines are marked in light green, stellar absorption lines are marked in orange, and interstellar absorption lines tracing neutral outflows are marked in dark green. } \label{fig:example_spec}
\end{figure*}

The unprecedented near-infrared sensitivity and wavelength coverage of JWST has driven major advances in our understanding of the role of outflows in quenching massive galaxies at cosmic noon. The quality of the data now available is illustrated in Fig. \ref{fig:example_spec}, which shows a deep JWST/NIRSpec spectrum of a massive galaxy at $z\sim$~2.4 observed using the G140M, G235M and G395M filters with integration times of 13h, 3.2h, and 1.6h, respectively. The galaxy exhibits weak emission lines and strong Balmer absorption lines, characteristic of a post-starburst system. Furthermore, the presence of deep, blueshifted \NaD\ absorption reveals a powerful neutral outflow. Thanks to JWST, we now have access to high signal-to-noise ratio spectra over the entire rest-frame optical wavelength range for hundreds of massive galaxies at cosmic noon, making it possible to measure the mass of neutral gas ejected by outflows, investigate their link to AGN activity and examine the impact of these outflows on galaxy star-formation histories. 

The vast majority of \NaD\ outflow studies at cosmic noon to date have been conducted using JWST/NIRSpec medium resolution spectra similar to what is shown in Figure~\ref{fig:example_spec}. These gratings provide a spectral resolution $R = c / \Delta v \simeq$~1000, corresponding to a velocity full width at half maximum of $\sim$~300~\kms. These observations can robustly identify outflows with bulk line-of-sight blueshifts $\gtrsim$~100~\kms, which translates to outflow velocities $\gtrsim$~200~\kms \citep[e.g.][]{Davies24NaD, Sun26}.

In this section, we summarize recent findings on the demographics and properties of neutral outflows in massive $z\sim$~2 galaxies (Section \ref{subsec:demographics}), their driving mechanisms (Section \ref{subsec:driving_mechanisms}), the role of outflows in quenching star-formation (Section \ref{subsec:impact}), and the long-term evolution of massive quiescent galaxies at cosmic noon (Section \ref{subsec:longterm_evolution}).

\subsection{Outflow demographics}\label{subsec:demographics}
The spectrum shown in Fig.~\ref{fig:example_spec} was obtained as part of the Blue Jay survey, a JWST Cycle 1 program (GO 1810) which targeted a mass-selected sample of 150 galaxies at 1.7~$<z<$~3.5 with the goal of examining the causes of galaxy quenching \citep{Belli25}. Here, we highlight results from the Blue Jay survey which was the first program to comprehensively examine the properties of neutral outflows at $z\sim$~2, but we note that the key findings have since been independently confirmed by several other JWST surveys, including EXCELS, DeepDive, and JADES \citep{Kehoe26, Sun26, Taylor26, Zhu26}. 

\begin{figure}
\centering
\includegraphics[scale=0.85, clip = True, trim = 15 55 0 0]{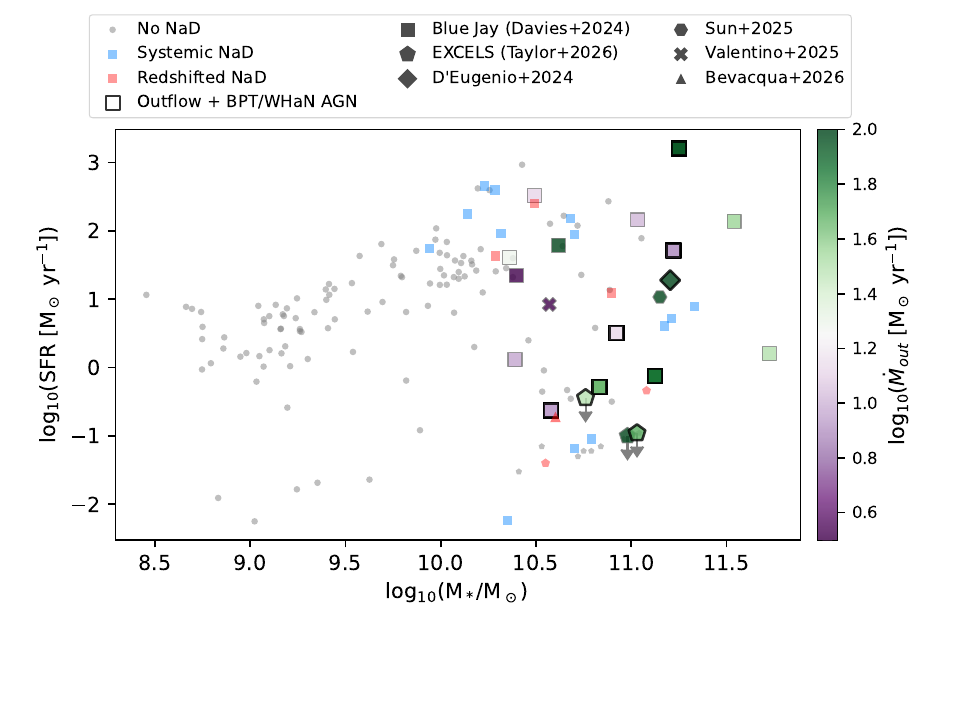}
\caption{Demographics of neutral outflows at 1.3 $ < z < $ 4.6 detected via \NaD\ absorption. Grey dots are galaxies from the Blue Jay sample with no \NaD\ absorption. Small blue and red markers indicate detections of systemic and redshifted \NaD\ absorption, respectively, from both Blue Jay and other samples (differentiated by the marker shape). Large green-to-purple markers indicate galaxies with \NaD\ outflows, color-coded by mass outflow rate. Black outlines denote outflow galaxies classified as AGN either using both the BPT and WHaN diagrams when [O~\textsc{iii}] and H$\beta$ are robustly measured, or using the WHaN diagram alone.}\label{fig:NaD_outflow_props}
\end{figure}

Strong \NaD\ absorption tracing neutral gas was detected in $\sim$~50\% of massive galaxies ($\log(M_*/M_\odot)>$~10) from the Blue Jay sample, spanning a range of star-formation rates from star-forming to quiescent systems (Fig.~\ref{fig:NaD_outflow_props}; \citealt{Davies24NaD}). Notably, the \NaD\ absorption detected in $z\sim$~2 quiescent galaxies is an order of magnitude stronger than what is seen in quenched systems at $z\sim$~0 and cannot be explained by excess absorption in stellar atmospheres, even when allowing for elevated sodium abundances. \NaD\ absorption is very rarely detected in lower mass galaxies, consistent with $z\sim$~0 results. Stacking analysis demonstrates that this is not due to weaker continuum flux in low mass galaxies \citep[e.g.][]{Davies24NaD} and may instead reflect the requirement for significant dust shielding to prevent sodium from being ionized \citep[e.g.][]{RobertsBorsani19}.

Within the Blue Jay sample, half of the \NaD\ absorption detections are blueshifted by at least 100~\kms\ relative to the galaxy systemic velocity, indicative of neutral outflows. A further 39\% are consistent with the galaxy systemic velocity and may trace outflows with lower (projected) velocities, or gas in the ISM. The existence of neutral outflows is further confirmed through the detection of blueshifted absorption in the Ca~\textsc{ii}~H and K lines, which are also resonant and thus trace cool neutral gas \citep{Liboni26}. The inferred neutral gas outflow rates ($\dot{M}_{\rm neutral}$) are at least as large as the galaxy SFRs, corresponding to mass loading factors $\eta = \dot{M}_{\rm neutral}/{\rm SFR} \gtrsim$~1, and the outflows have velocities of \mbox{200~--~1000~\kms}, lower than or comparable to the galaxy escape velocities \citep{Davies24NaD}. 

When both neutral (\NaD\ absorption) and ionized gas (optical emission line) outflows are detected, the neutral phase dominates the outflow mass budget but travels more slowly than the ionized gas \citep[e.g.][]{Belli24, DEugenio24, Bugiani25, Wu25}, consistent with the canonical picture of multiphase winds developed from detailed studies of nearby outflows \citep[e.g.][]{Leroy15}. Overall, these results suggest that the majority of massive galaxies at cosmic noon host powerful neutral outflows that rapidly eject large amounts of cool gas from galaxies but likely do not permanently remove this material from the host halos.

\subsection{Outflow driving mechanisms}\label{subsec:driving_mechanisms}
In galaxies with little current star-formation, several lines of evidence suggest that the observed outflows are primarily driven by processes linked to supermassive black hole accretion such as radiation pressure, accretion disk winds, or radio jets. Many quiescent galaxies show high-ionization emission lines characteristic of radiative AGN activity (e.g. elevated [O~\textsc{iii}]/H$\beta$ or [Ne~\textsc{iii}]/[O~\textsc{ii}] ratios or [Ne~\textsc{v}]$\lambda$3427 emission; \citealt{Davies24NaD, Belli24, DEugenio24, Bugiani25, Wu25, Sun26, Taylor26, Valentino26}). In other cases, the spectra are dominated by strong low-ionization emission lines (e.g. elevated [N~\textsc{ii}]/H$\alpha$ and [S~\textsc{ii}]/H$\alpha$ ratios; \citealt{Bugiani25, Sun26, Taylor26}) which could trace either low luminosity AGN or shock excitation by radio jets \citep[e.g.][]{Davis12, Ilha22, Couto23}. We note that $>$~50\% of all [N~\textsc{ii}]/H$\alpha$-selected AGN at cosmic noon do not show AGN signatures at X-ray or radio wavelengths, likely because the available observations are not sensitive to moderate luminosity AGN at these distances \citep[e.g.][]{NMFS19}. Ionized outflows from galaxies with low levels of star-formation often reach velocities of 800~--~1000~\kms\ \citep[e.g.][]{Belli24, DEugenio24, Bugiani25}, consistent with typical AGN-driven outflows and substantially higher than typical star-formation-driven outflows at $z\sim$~2 (see discussion in Section \ref{subsec:cosmic_noon}). In most cases, the rates of energy and/or momentum injection by supernovae are insufficient to power the observed outflows, whereas the detected AGN are able to supply the required energy \citep[e.g.][]{Davies24NaD, Sun26}. 

It is important to consider whether the observed \NaD\ absorption in quiescent galaxies could trace `fossil' outflows, meaning outflows that were launched by past star-formation or AGN activity that has since faded. This is especially relevant for post-starburst galaxies which have experienced a recent episode of strong star formation that could be responsible for the observed outflow. However, it is unclear how long outflows remain observable once the driving source fades. Previous studies suggest that starburst-driven outflows can persist for a maximum of 25~Myr \citep{McQuinn18}, significantly shorter than the typical post-burst timescales which are hundreds of Myr. AGN flicker on much shorter (Myr) timescales, and hydrodynamical simulations of outflows in idealized galaxy bulges predict that in gas-poor galaxies, fossil outflows should outnumber outflows powered by current AGN activity \citep{Zubovas23}. Consistent with this picture, there are multiple examples of quiescent galaxies that show powerful outflows but little or no evidence for optical (radiatively efficient) AGN activity, potentially revealing the signatures of past AGN episodes \citep[e.g.][]{Almaini25, Sun26a, Sun26, Taylor26}. It is also possible, however, that these outflows are driven by AGN while in a low-luminosity (or radiatively inefficient) state \citep[e.g.][]{Bugiani25, Hewitt26}. A key unknown in determining whether fossil outflow scenarios are plausible is how long ago the outflows were launched. To calculate this, we need to know how far the outflows have traveled from their point of launch. This is very poorly constrained, with current estimates ranging from 1~kpc to 40~kpc (see Section~\ref{sec:open_questions}). A 200~\kms\ outflow 1~kpc from the galaxy would have a travel time of 5~Myr, meaning that the current SFR should be similar to the SFR when the outflow was launched. The same outflow at a distance of 40~kpc would have a travel time of $\sim200$~Myr; comparable to observed post-burst timescales and making a fossil star-formation-driven outflow scenario more plausible. 

For main sequence galaxies, it is more difficult to conclusively determine the outflow driving mechanism. Initial indications suggest that galaxies with \NaD-detected outflows have a higher AGN incidence than the underlying massive galaxy population. However, the ongoing star-formation activity provides sufficient energy and momentum to power the observed outflows, and there is tentative evidence that the outflow velocity and mass outflow rate are weakly correlated with the level of star-formation activity  \citep[e.g.][]{Davies24NaD, Sun26}. Measuring the ionized outflow velocities in these systems would help to distinguish between AGN-driven and SF-driven winds, which have typical ionized outflow velocities of \mbox{1000~--~2000~\kms} and 500~\kms, respectively \citep[e.g.][]{NMFS19}. 

\subsection{Role of outflows in quenching massive galaxies}\label{subsec:impact}
Neutral outflows are observed in 25~--~50\% of galaxies with $\log(M_*/M_\odot)>$~10 at cosmic noon (see Section \ref{subsec:demographics}). To understand how these outflows impact galaxy evolution and star-formation quenching, it is important to examine their link with star-formation histories. There are now several known examples of massive, quenched \mbox{$z\sim$~1~--~4} galaxies that experienced a rapid decline in star-formation activity within the last \mbox{300~Myr~--~1~Gyr} and show powerful neutral outflows with $\eta \gtrsim $~1~--~10, sufficient to quench star-formation \citep[e.g.][]{Belli24, DEugenio24, Wu25, Sun26a}. These case studies provide the most direct observational evidence thus far that gas removal by powerful AGN-driven ejective feedback drives rapid star-formation quenching in massive galaxies at cosmic noon. 

Statistical studies of quiescent galaxies suggest a tentative link between post-quenching timescale and outflow properties. Within the $z\sim$~2 Blue Jay sample, neutral outflows are detected in 75\% of galaxies that quenched within the last 500~Myr, whereas the neutral outflow fraction is 0\% for more evolved systems \citep{Park24}. Similarly, all ionized outflow detections are in galaxies $<$~500~Myr post-quenching \citep{Bugiani25}. In the slightly higher redshift EXCELS and \textit{DeepDive} samples ($z\sim$~3~--~4), 23~--~30\% of galaxies $\leq$~500~Myr post-quenching show neutral outflows \citep{Taylor26, Zhu26}, whereas older galaxies at the same redshift do not show neutral outflows \citep[e.g.][]{Forrest26}. In $z\sim$~1 post-starburst galaxies, \MgII\ outflows are detected up to 1~Gyr after the last burst \citep{Taylor24}. These findings are broadly consistent with studies of starburst, post-starburst and quiescent galaxies at $z\sim$~0 which have shown that the incidence and velocity of \NaD\ outflows decline as galaxies age and quench \citep[e.g.][]{Sun24}. \citet{Park24} found that the oldest quenched galaxies at $z\sim$~2 show weak absorption consistent with the galaxy systemic velocity, perhaps tracing older, lower-velocity outflows that are undetectable in the R~$\sim$~1000 JWST observations.

These observational results establish powerful outflows as a compelling candidate to explain rapid quenching of massive galaxies at cosmic noon. Previous studies have found evidence for two quenching pathways, defined by their characteristic timescales: fast quenching \mbox{($\sim$~100~--~300~Myr)} and slow quenching ($\sim$~1~Gyr) \citep[e.g.][]{Wu18, Belli19, Carnall19, Suess21, Park23}. Slow quenching is often attributed to starvation of gas supply and depletion of remaining gas by star-formation \citep[e.g.][]{Larson80, Peng15, Spitoni17, Trussler20}. On the other hand, fast quenching is hypothesised to be triggered by dramatic events like mergers or violent disk instabilities which funnel gas towards the centres of galaxies, trigger centrally concentrated starbursts and AGN accretion, and drive powerful outflows \citep[e.g.][]{Hopkins06, Barro13, Dekel14, Zolotov15, Tacchella16}. There is some evidence that rapid star-formation quenching at cosmic noon is preceded by bursts of star-formation \citep[e.g.][]{Park23, Belli24, DEugenio24, Sun26a}. However, only the strongest bursts can be recovered from the observed spectroscopy in a way that is robust to the choice of star-formation history prior \citep[e.g.][]{Park24}. If outflow-driven quenching is triggered by galaxy compaction events, then we would also expect the strongest outflows to be found in starburst galaxies, as is the case in the local Universe \citep[e.g.][]{Sun24}. Starburst galaxies are rare and not well represented in current samples \citep[e.g.][]{Davies24NaD, Sun26}. There are tentative indications that galaxies with higher specific SFRs may drive faster and more energetic outflows \citep[e.g.][]{Sun26}, but the uncertainties on individual measurements are large due to many factors including low velocity resolution, unknown outflow geometry and uncertainties in converting \NaD\ absorption measurements to hydrogen column densities (see Section \ref{sec:open_questions}). Observations of larger galaxy samples at higher spectral resolution are needed to determine the relationship between bursts of star-formation and strong outflows at cosmic noon.

\subsection{Long-term evolution of massive quiescent galaxies}\label{subsec:longterm_evolution}

While there is compelling evidence that powerful ejective feedback can trigger rapid quenching, it is less clear how these quenched galaxies evolve over longer timescales. \citet{Davies24NaD} find that 71\% of neutral outflows at $z\sim$~2 are too slow to escape the host halo, in contrast with $z\sim$~2 AGN-driven ionized outflows which typically exceed the galaxy escape velocity \citep[e.g.][]{NMFS19, Leung19, Belli24, DEugenio24, Bugiani25}. This is consistent with predictions from ISM simulations that cool gas entrained in outflows moves significantly slower than the surrounding hot wind \citep[e.g.][]{Gronke18, Kim20, Vijayan20, Ward24}. Given the low velocities, the outflowing material may eventually be re-accreted in a galactic fountain \citep[see also][]{Sun26, Zhu26}. Indeed, \citet{Bevacqua26} discovered a strong \NaD-detected neutral inflow feeding a massive galaxy at $z$~=~2.7 that has nevertheless been quiescent for the past 1 Gyr, and there are several other examples of quiescent galaxies with detected \NaD\ inflows \citep[e.g.][]{Davies24NaD, Sun26, Taylor26}. Measurements of neutral gas column density and velocity as a function of galactocentric distance along multiple sightlines for individual galaxies would provide crucial insights into the fate of outflowing neutral gas.

Furthermore, it remains unclear whether powerful outflows completely remove all the cold gas from galaxies. The weak systemic absorption detected in galaxies $>$~500~Myr post-quenching suggests that these galaxies may either still have, or recently had, significant amounts of neutral gas in their ISM. \citet{Taylor26} found no evidence for systemic absorption in $z\sim$~3 quenched galaxies from EXCELS; however, all of their galaxies are observed less than 500~Myr after quenching when outflows are more dominant. \citet{Scholtz26} measured a molecular gas fraction of $<$~1\% in a $z\sim$~3 massive quiescent galaxy with a strong neutral outflow, and showed that star-formation could plausibly have consumed all of the galaxy's gas, with outflows balancing inflows to prevent replenishment of the gas reservoir. This is only one possible evolutionary history, but demonstrates the potential power of combining star-formation history, outflow and molecular gas measurements for large samples of cosmic noon galaxies at different quenching stages. 

\begin{figure}
    \centering
    \includegraphics[width=\linewidth]{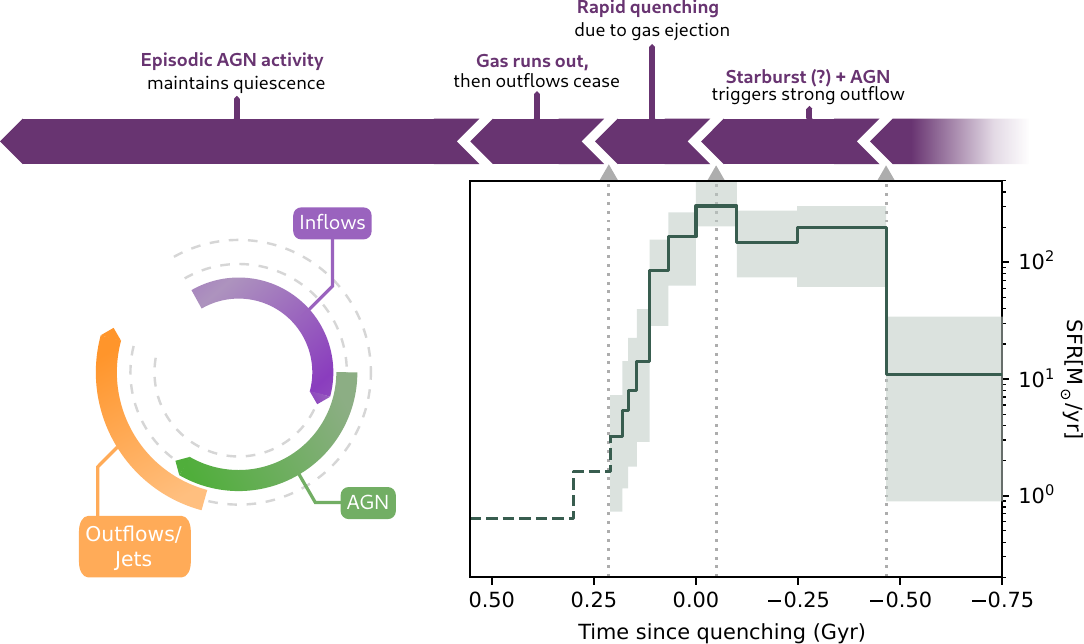}
    \vspace{-10pt}
    \caption{Schematic summarising the possible mechanisms driving the rapid quenching and ongoing quiescence of massive galaxies at cosmic noon. On the right, the solid green line and shaded region show the star-formation history of COSMOS-11142 \citep{Belli24}, and the dashed line shows a possible extrapolation into the future for illustration purposes. Strong gas inflow/accretion likely triggered a burst of star-formation and AGN activity, driving a powerful outflow. This outflow ejected a large amount of cold gas, leading to rapid quenching and leaving the galaxy below the main sequence where it continues to drive a strong outflow \citep{Belli24}. The left side of the figure illustrates what may happen to COSMOS-11142 in the future. As the gas runs out, the galaxy may enter a phase of episodic AGN activity, where gas inflows trigger AGN activity that then powers outflows or radio jets, preventing any new or remaining gas from forming stars \citep[e.g.][]{Park24, Almaini25, Taylor26}. }
    \label{fig:quenching_timeline}
\end{figure}

AGN heating could be a key mechanism preventing galaxies from rejuvenating. Current estimates suggest that between 20~--~50\% of passive galaxies at \mbox{1 $<z<$ 5} display optical emission line properties consistent with weak radiative AGN activity or shocks produced by radio jets interacting with the ISM \citep[e.g.][]{Slob24, Bugiani25, Skarbinski26, Stevenson26}. At $z\sim$~0, old quiescent galaxies with neutral gas inflows show an increased incidence of radio jets \citep[e.g.][]{Roy21, Moghni26}. Post-starburst galaxies at cosmic noon have relatively weak and compact radio jets, whereas quiescent galaxies have higher radio luminosities and more extended morphologies, suggesting that maintenance mode feedback becomes stronger as galaxies age \citep[e.g.][]{Patil26}. 

Another possible source of long-timescale feedback is episodic weak outflows. Quenched galaxies may go through cycles of fueling and feedback, whereby inflows fuel the nuclear region, leading to increased AGN activity and outflows which remove cold gas, causing the AGN to turn off until inflows build up enough gas to trigger another AGN episode \citep[e.g.][]{Almaini25, Taylor26}. Consistent with this picture, \citet{French25} identified four local post-starburst galaxies with significant recent brightening in radio emission, and demonstrated that this newly injected energy is sufficient to drive outflows and turbulence capable of suppressing star-formation in these systems. This cycling could continue until the inflows cease or become too weak to fuel AGN activity. Observations of more quiescent galaxies over a wider range of ages are needed to understand how long episodic AGN and outflows persist in these systems.

Fig.~\ref{fig:quenching_timeline} summarises the possible mechanisms driving the rapid quenching and ongoing quiescence of massive galaxies at cosmic noon.

\section{Testing Feedback Models with Multiphase Outflow Measurements}
The availability of neutral outflow measurements for large samples of massive $z\sim$~2 galaxies presents a unique opportunity to study multiphase outflow properties and test feedback models used in cosmological simulations. Feedback models are tuned to match galaxy population statistics such as the galaxy stellar mass function and the cosmic SFR density (see e.g. \citealt{Somerville15, Naab17, Crain23} and references therein), but strong degeneracies between model parameters mean that simulations employing vastly different combinations of ejective and preventative feedback can produce galaxies with roughly the same stellar properties \citep[e.g.][]{Crain23, Wright24}. Directly measuring outflow properties provides a means to break this degeneracy.

One of the most fundamental observables for testing feedback models is the relationship between galaxy stellar mass ($M_*$) and outflow mass loading factor ($\eta$). The $M_*-\eta$ relation is set by a combination of stellar feedback at lower masses and AGN feedback at higher masses \citep[e.g.][]{Nelson19, Mitchell20} and thus constrains several key aspects of feedback models. In most cosmological simulations, AGN feedback becomes dominant at $\log(M_*/M_\odot) \gtrsim$~10.5~--~10.7, primarily because the feedback models are calibrated to reproduce the knee in the galaxy stellar mass function \citep[e.g.][]{Crain15, Weinberger17, Dave19, Dolag25}. Directly measuring the mass at which the $M_*-\eta$ scaling changes would reveal whether the knee in the galaxy stellar mass function is indeed primarily caused by a change in the dominant feedback mode. The slope and normalization of the $M_*-\eta$ relation above the transition mass constrains the relative importance of ejective and preventative AGN feedback: simulations with dominant ejective feedback predict that $\eta$ should increase with mass, whereas those with dominant preventative feedback predict flatter scalings and/or lower normalization \citep[e.g.][]{Bower17, Zinger20}. Similarly, if low mass galaxies have significantly weaker outflows than predicted by simulations, this may indicate that the star formation is being suppressed through heating rather than direct gas ejection \citep[e.g.][]{Smith24, Bennett25}.

Many observational studies have presented measurements of $\eta$ as a function of $M_*$ \citep[e.g.][]{NMFS19, Fluetsch19, Llerena23, Marasco23, Carniani24, Weldon24}. Recent compilations of multiple samples suggest that outflows from low mass galaxies are weaker than predicted by simulations, although there is significant scatter between individual measurements \citep[e.g.][]{MazzilliCiraulo25, Davies26, HernandezYevenes26}. However, the vast majority of these studies are based on ionized gas alone. Small samples of outflows measured in both ionized and cool (atomic/molecular) phases suggest that including the cool component boosts $\eta$ by about an order of magnitude \citep[e.g.][]{Fluetsch19, Davies26, HernandezYevenes26}. Adding the cool component is also likely to change the slope of the measured $M_*-\eta$ relation. As discussed in Section \ref{subsec:multiphase_properties}, the amount of cool gas in simulated outflows depends sensitively on the balance between cloud destruction, radiative cooling, and turbulent mixing between outflow phases \citep[e.g.][]{Gronke18, Ji19, Fielding20, Chen23, Ward24}. ISM simulations predict that the ionized-to-total and molecular-to-atomic ratios vary significantly as a function of star-formation rate surface density (which sets the strength of the incident radiation field; \citealt{Rathjen23}) and gas density (which regulates the self-shielding in the cold gas; \citealt{Vijayan24b}). Measuring the fraction of outflowing mass in different gas phases as a function of galaxy properties would provide valuable constraints on physical models of cold gas in galactic winds.

In summary, multiphase measurements of $\eta$ for large samples of galaxies at $\log(M_*/M_\odot)>$~10 would provide key constraints on the fraction of cold gas in galactic winds, the stellar mass where AGN feedback becomes dominant, and the strength of ejective feedback, especially in high mass galaxies. Measuring the cool gas outflow rates from lower mass galaxies remains challenging due to a lack of robust outflow mass tracers in these systems (see Fig.~\ref{fig:tracers}).

\section{Uncertainties on mass outflow rates}\label{sec:open_questions}
JWST has provided a wealth of insights into the properties of neutral outflows and their impact on star-formation, but outflow rate measurements are still subject to significant systematic uncertainties. Computing neutral mass outflow rates from measurements of blueshifted sodium absorption requires assumptions about 1) the sodium abundance, neutral fraction and dust depletion (to convert the gas-phase neutral sodium column density to a hydrogen column density), and 2) the geometry of the outflowing material (to convert column density to total mass and outflow rate). Appendix~\ref{secA1} outlines the key equations used in these calculations. 

There are virtually no constraints on the abundance, neutral fraction or dust depletion of sodium in galactic winds, and most studies adopt canonical values measured for cold clouds in the Milky Way \citep{Rupke05a, Shih10, Baron22}. These values may not be representative of the conditions in outflows, and there is likely to be significant object-to-object variation. Even within the Milky Way, dust depletion of Na varies by 0.5~dex \citep{Wakker00}. Dust depletion generally increases in denser gas, but dust may be destroyed in outflows due to interaction with the hot wind fluid \citep[e.g.][]{Farber22, Richie26}. Similarly, the neutral fraction in outflows may vary by more than an order of magnitude due to differences in radiation field strength and gas density \citep[e.g.][]{Murray07, Baron20}. Constraints on sodium-to-hydrogen ratios in extragalactic environments can be obtained using absorption systems against bright background continuum sources detected in both Na~\textsc{i} and H~\textsc{i}. \citet{Moretti26} detected \NaD\ and Ly$\alpha$ absorption in a Damped Lyman Alpha system at $z$~=~2.4 and measured a N(H~\textsc{i})/N(Na~\textsc{i}) ratio 30\% lower than the canonical value, highlighting the significant systematic uncertainties on individual outflow mass measurements. In the coming years, large surveys will significantly increase the number of extragalactic absorption systems with detections of \NaD\ combined with either Ly$\alpha$ (e.g. DESI; \citealt{Weng26}; 4MOST; \citealt{Peroux23}) or H~\textsc{i}~21cm absorption (e.g. ASKAP-FLASH; \citealt{Weng22}). These statistical samples will allow us to quantify for the first time how the N(Na~\textsc{i})-to-N(H~\textsc{i}) conversion factor varies with galaxy properties such as metallicity and dust attenuation. When direct N(H~\textsc{i}) measurements are not available, measurements of other low-ionization resonant absorption lines (e.g. Ca~\textsc{ii}, \MgII, \FeII) can help determine the true hydrogen column density \citep[e.g.][]{Valentino25, Wu25, Liboni26}.

The geometry of neutral outflows is also poorly constrained. Most \NaD\ outflow measurements beyond the local Universe come from single aperture spectra. 
Resolved studies at $z\sim$~0 find that \NaD\ absorption is typically confined within $\lesssim$~10~kpc of galaxy nuclei \citep[e.g.][]{Martin06, Rupke15, Rupke17, Baron20, RobertsBorsani20, Avery22, Rubin22}; likely due to the requirement for significant dust shielding.
\citet{DEugenio24} used the JWST/NIRSpec IFU to map a \NaD\ outflow across a $z\sim$~3 post-starburst galaxy and measured an extent of at least 2.7~kpc, consistent with the $z\sim$~0 results. \citet{Belli24} measured outflowing ionized gas extending 3~kpc from the centre of a galaxy that also shows a strong neutral outflow. \MgII\ outflows at $z\sim$~0.7~--~2.0 are similarly found to have a typical spatial extent of $\sim$~5~kpc \citep{Pessa26}. However, outflows are most easily mapped in face-on galaxies where their size traces the projected extent across the galaxy disk, rather than distance from the galaxy which is the key quantity for outflow rate measurements. Again, quasar absorption line measurements provide complementary constraints on outflow sizes. \citet{Moretti26} detected neutral sodium 38~kpc from a quenched galaxy at $z\sim$~2.4 using quasar absorption line spectroscopy, suggesting that the sizes of \NaD\ outflows may vary by up to an order of magnitude. The results presented in \citet{Davies24NaD} and \citet{Sun26} assume $r_{\rm out}$~=~1~kpc and thus represent a conservative lower limit on the mass outflow rates.

The solid angle $\Omega$ can be indirectly constrained through multiple channels. Using stacked spectra from the Sloan Digital Sky Survey, \citet{Chen10} showed that face-on galaxies display more blueshifted \NaD\ absorption than edge-on galaxies which show primarily systemic absorption, consistent with expectations for non-spherical ($\Omega <4\pi$) outflows oriented perpendicular to the galaxy disk. The fraction of galaxies with detected outflows $f_{\rm out}$ is determined by the product of the outflow covering fraction ($C_\Omega=\Omega/4\pi$) and the outflow duty cycle \citep[e.g.][]{Xu25}. Since the duty cycle must be $\leq$~1, this implies $C_\Omega \geq f_{\rm out}$. For a segment of a sphere subtending a solid angle $\Omega$, the half-opening angle $\theta$ is given by $\Omega = 4\pi(1-{\rm cos}(\theta))$, and thus $f_{\rm cov}=1-{\rm cos}(\theta)$. At $z\sim$~2, \NaD\ outflows are detected in 25~--~50\% of massive galaxies \citep[e.g.][]{Davies24NaD, Sun26, Taylor26}, implying wide-angle winds with half opening angles of $\theta \geq$~40~--~60$^\circ$. This is consistent with the wide opening angles measured for local starburst-driven superwinds \citep[see][and references therein]{Veilleux20} as well as outflows observed in \MgII\ emission at cosmic noon \citep{Pessa26}, and similar to or larger than opening angles measured in local Seyfert galaxies \citep[e.g.][]{Revalski21}. However, outflow studies using rest-frame UV absorption lines at $z\gtrsim$~2 have reported correlations between outflow velocity and galaxy inclination, consistent with collimated outflows \citep[e.g.][]{Kehoe26}. Better constraints on outflow geometry will require spatially resolved mapping of \NaD\ outflows. Mapping \NaD\ outflows at lower redshifts, and especially in systems where extraplanar \NaD\ is detected directly in emission, will provide valuable constraints on the typical geometries of these winds.

\section{Summary and outlook}\label{sec:summary}
JWST has transformed our understanding of outflows and their role in galaxy quenching. In particular, the ability to detect neutral gas outflows and link them with galaxy star-formation histories and AGN activity has revealed the strongest observational evidence thus far for a direct link between AGN-driven ejective feedback and star-formation quenching in massive galaxies at cosmic noon. 

These results may also provide important clues about how the earliest massive quiescent galaxies formed. The star-formation histories of massive quiescent galaxies at $z\sim$~3~--~4 show prominent peaks of star-formation followed by rapid quenching \citep[e.g.,][]{Carnall23, Carnall24, Glazebrook24, Nanayakkara25, DeGraaff25, Edward26}, similar to what is observed at cosmic noon. Simulations predict that powerful AGN-driven outflows likely play a key role in the formation of the first massive quiescent galaxies \citep[e.g.][]{Kimmig25, Lagos25, Szpila25, ChandroGomez26}. 
As massive quiescent galaxies are discovered at increasingly high redshift \citep{Weibel25}, it is crucial to investigate the role of AGN-driven outflows in the rapid cessation of star formation in these systems. Observations of \NaD, \FeII\ and \MgII\ absorption lines have revealed the presence of neutral outflows in two quiescent galaxies at $z\sim4$ and $z\sim7$ \citep{Wu25, Valentino25}, providing further evidence for AGN-driven ejection of gas leading to rapid quenching. 

Additional insights into the formation of early massive quiescent galaxies can be gained by studying their likely progenitors. The inferred high star-formation rates prior to quenching suggest that these galaxies may be descended from $z\sim$~5 starbursts or sub-millimetre galaxies \citep[e.g.][]{Toft14, Valentino20, ArayaAraya26}. Studies based on ALMA and JWST have shown that stellar feedback in such systems can drive very powerful outflows that may exhaust their molecular gas reservoirs without the need for AGN feedback \citep[e.g.][]{Spilker18, Spilker20, Butler23, Davies26}. Future surveys of outflows in early massive quiescent galaxies and their progenitors would provide unprecedented insights into the physical processes leading to rapid quenching in the early Universe.

\backmatter

% \bmhead{Supplementary information}

% If your article has accompanying supplementary file/s please state so here. 

% Authors reporting data from electrophoretic gels and blots should supply the full unprocessed scans for key as part of their Supplementary information. This may be requested by the editorial team/s if it is missing.

% Please refer to Journal-level guidance for any specific requirements.

\bmhead{Acknowledgements}
We thank the referee for their valuable feedback on this paper. RLD is supported by the Australian Research Council through the Discovery Early Career Researcher Award (DECRA) Fellowship DE240100136 funded by the Australian Government. SB is supported by ERC grant 101076080 ``Red Cardinal''. This research was supported by the International Space Science Institute (ISSI) in Bern, through ISSI International Team project 24-602 ``Multiphase Outflows in Galaxies at Cosmic Noon''. 

\begin{appendices}

\section{Computing Mass Outflow Rates from Na~I~D Absorption}\label{secA1}
The first step in measuring the neutral mass outflow rate from \NaD\ absorption is to fit and divide out the stellar continuum, including stellar \NaD\ absorption, using a stellar population fitting code such as \textsc{Prospector} \citep{Leja19, Johnson21} or \textsc{ppxf} \citep{Cappellari23}. This leaves the interstellar \NaD\ absorption profile as a function of velocity, $f_{\rm Na~I~D}(v)$. The insterstellar \NaD\ profile is often parametrized using the partial covering model \citep{Rupke05b} which constrains the optical depth ($\tau$) and covering fraction ($C_f$) of the absorbing gas against the background continuum source:
\begin{equation}
f_{\rm Na~I~D}(v) = 1 - C_f + C_f \exp \left(-\tau_b(v) - \tau_r(v) \right)
\end{equation}
Here, $\tau_b(v)$ and $\tau_r(v)$ are the optical depth profiles of the \NaD\ 5891\AA\ and 5897\AA\ lines, respectively. Most studies adopt either a Gaussian or Voigt profile for $\tau(v)$. The optical depth ratio of the line centers is fixed at $\tau_{b,0}$~=~2~$\tau_{r,0}$, reflecting the known doublet ratio \citep{Draine11}. Note that this is different from the ratio of the absorption equivalent widths, which varies from 2 in the optically thin regime to 1 in the optically thick regime. The optical depth at the centre of the red line ($\tau_{r,0}$) can then be converted to Na~\textsc{i} column density following \citet{Draine11}:
\begin{equation}
{\rm N_{Na~I}(cm^{-2})} = 10^{13} \left(\frac{\tau_{r,0}}{0.7580}\right) \left(\frac{0.4164}{f_{\rm lu}}\right)\left(\frac{\rm 1215\AA}{\lambda_{\rm lu}}\right)\left(\frac{b}{\rm 10 \, km \, s^{-1}}\right)
\end{equation}
Here $f_{\rm lu}$ = 0.32 is the oscillator strength, $\lambda_{\rm lu}$ = 5897\AA\ is the rest-frame wavelength of the transition, and $b$ is the Doppler parameter, related to the Gaussian velocity dispersion $\sigma$ by $b = \sqrt{2} \sigma$. The Na~\textsc{i} column density can then be converted to a total gas column density following \citet{Shih10} and \citet{Baron22}:
\begin{equation}
    {\rm N_H(cm^{-2})} = \frac{\rm N_{Na~I}(cm^{-2})}{(1-y) \, 10^{A+B+C}}\label{eqn:nh}
\end{equation}
Here ($1-y$) is the sodium neutral fraction, $A$ is the solar sodium abundance, $B$ accounts for the depletion of sodium onto dust grains, and $C$ is the metallicity relative to solar. Most studies adopt canonical values of \mbox{$B$ = $-0.95$} \citep{Savage96} and $1-y$~=~0.1 \citep{Rupke05a} based on measurements of cool diffuse clouds in the Milky Way. 

To convert the line-of-sight gas column density to a mass outflow rate, one must account for the outflow geometry. Unresolved studies of \NaD\ outflows typically adopt the classical `thin-shell' model \citep[e.g.][]{Heckman00, Martin05, Rupke05b}, which assumes a spherically symmetric mass-conserving shell of outflowing material at a radius $r_{\rm out}$ that covers a fraction of the solid sphere given by $C_\Omega$. The inferred mass outflow rate scales linearly with the size of the outflow shell which in turn scales linearly with $r_{\rm out}$ and $C_\Omega$. The small-scale covering fraction of the absorbing material against the background continuum source, $C_f$, accounts for clumpiness of the outflowing gas. Under these assumptions, the mass outflow rate is given by 
\begin{equation}
    \dot{M}_{\rm out}({\rm M_\odot\, yr^{-1}}) = 11.45 \, \left(C_f \frac{C_\Omega}{0.4}\right) \left(\frac{\rm N_H}{10^{21} \, \rm cm^{-2}}\right) \frac{r_{\rm out}}{1 \, \rm kpc} \frac{v_{\rm out}}{200 \, \rm km/s}\label{eqn:mdot}
\end{equation}

The outflow size and velocity can also be combined to estimate the dynamical time $t_{\rm dyn}$:
\begin{equation}
t_{\rm dyn} ({\rm Myr}) \simeq 5 \, \frac{r_{\rm out}}{1 {\rm \, kpc}}\frac{200~\kms}{v_{\rm out}} \label{eqn:tdyn}
\end{equation}

% An appendix contains supplementary information that is not an essential part of the text itself but which may be helpful in providing a more comprehensive understanding of the research problem or it is information that is too cumbersome to be included in the body of the paper.

% %%=============================================%%
% %% For submissions to Nature Portfolio Journals %%
% %% please use the heading ``Extended Data''.   %%
% %%=============================================%%

% %%=============================================================%%
% %% Sample for another appendix section			       %%
% %%=============================================================%%

% %% \section{Example of another appendix section}\label{secA2}%
% %% Appendices may be used for helpful, supporting or essential material that would otherwise 
% %% clutter, break up or be distracting to the text. Appendices can consist of sections, figures, 
% %% tables and equations etc.

\end{appendices}


\bibliography{mybib}% common bib file

\end{document}